\documentclass[prb,preprint]{revtex4-1}

\usepackage{amsmath}
\usepackage{amsfonts}
\usepackage{graphicx}
\usepackage{color}
\usepackage{ulem}
\usepackage{cancel}

\newcommand{\blue}[1]{\textcolor{blue}{#1}}
\renewcommand{\blue}[1]{\textcolor{black}{#1}} 
\newcommand{\red}[1]{\textcolor{red}{#1}}
\renewcommand{\red}[1]{} 

\begin{document}

\title{Feynman, Lewin, and Einstein Download Zoom: A Guide for Incorporating E-Teaching of Physics in a Post-COVID World}

\author{Daniel J. O'Brien}
\email{djo44@georgetown.edu}

\affiliation{Department of Physics, Georgetown University, Washington, DC 20057}

\date{\today}

\begin{abstract}
Distance education has expanded significantly over the last decade, but the natural sciences have lagged in the implementation of this instructional mode. \blue{T}he abrupt onset of the COVID-19 pandemic left educational institutions scrambling to adapt curricula to distance modalities. With projected effects lasting through the 2020--21 academic year, this problem will not go away soon. Analysis of the literature has elucidated the costs and benefits of, as well as obstacles to, the implementation of e-learning, with a focus \blue{on} undergraduate physics education. Physics faculty report that a lack of time \blue{to learn} about research-driven innovation is their primary \blue{barrier} to implementing it. In response, this paper is intended to help physics lecturers and lab instructors re-think their courses \blue{now that distance learning is far more prevalent due to the pandemic}.  This paper serves as an all-in-one guide of recommendations \blue{for} successful distanced educational practices, with an emphasis on smartphones and social media. These technologies were chosen for their utility in a virtual environment. Additionally, this paper can \blue{be used}  as a resource for university administrators to adapt to the changing needs associated with new teaching modalities.  
\end{abstract}

\maketitle 

\section{Introduction} 
Despite much debate, no consensus has been formed in the literature as to a universal definition of e-learning. \cite{Singh2019, Moore2011, GuriRosenblit2005} For the purposes of this paper, it is defined as ``technology-based learning in which learning materials are delivered electronically to remote learners via a computer network.'' \cite{Zhang2004} E-learning can be divided into two categories: asynchronous and synchronous. The former is commonly implemented through a combination of pre-recorded videos, email, and discussion boards. The latter is usually implemented through a combination of videoconferencing and chat platforms (Zoom, WebEx, Skype, etc.). \cite{Hrastinski2008} Within the literature, ``virtual'' learning often refers to synchronous methods, whereas ``online'' refers to asynchronous. 

From 2002--2016, distance enrollment at higher education institutions rose dramatically, averaging an increase of 18.5\% per year, largely driven by e-learning. Meanwhile, on-campus enrollment dropped by 6.4\% between 2012--2016. \cite{Seaman2018, Allen2003} When compared to students in other fields, undergraduate physical science students consistently rank near the bottom in terms of the ratio of online to in-person classes taken. In the 2015--16 academic year, for example, although 43.1\% of the entire U.S. undergraduate population took an online course, only 32.8\% of students in the physical sciences did so.\cite{NCES2018} However, a few universities have housed online physics classes for decades, demonstrating the field's ability to thrive over time. \cite{Kortemeyer2017, Lambourne2007, Lambourne2005, Adams2003}

The COVID-19 pandemic profoundly impacted academic institutions, causing most US universities to shut down on-campus classes and ousting students from their dormitories before the scheduled end of the 2019--20 school year. In an attempt to maintain instructional continuity, teachers turned to videoconferencing and recordings of lectures, labs, and office hour sessions. This change may be particularly detrimental to students within STEM subjects due to at-home students' lack of access to instructional technologies critical to STEM learning. \cite{Sintema2020} 

This pandemic may therefore act as a motivator for physics faculty and administrators to update curricula, adopt novel teaching modalities, and embrace research-based innovations. This change is not unreasonable; a survey of U.S. physics faculty found that 92\% reported that their department encouraged improving instruction. Nearly half (48\%) of the surveyed faculty reported that they currently use at least one research-based innovation strategy in their teaching. Unfortunately, 53\% of those answering replied that the principal reason for not using more research-driven innovations in their classroom is a lack of time (especially time to research and implement changes).\blue{\cite{Dancy2010, Henderson2007}} As a response, the aim of this manuscript is to act as a brief but thorough guide for educators. \blue{This article presents} an explanation for some of the above trends, \blue{as well as} specific guidance for implementing techniques and \blue{e-learning} systems in physics. 

First, the benefits of, barriers to, and key factors for the implementation of physics e-learning will be discussed. Next, \blue{the fact that e-learning has an unequal impact on students from different demographic groups (based on gender, household income) is addressed}. Following that, the smartphone will be introduced as an important educational tool for physics e-learning. The smartphone can be very useful for both doing science (data collection, analysis, demonstrations) as well as facilitating other technologies, such as social media. Applications of social media in the traditional and virtual classrooms are then presented, and their use is examined in depth. Other technologies for e-learning are then briefly introduced. Finally, the need for extensive institutional support to facilitate e-learning has been widely recognized and thoroughly researched. Consequently, a guide for administrators on the \blue{keys to} success in this implementation is presented. In all, it is the goal of this manuscript to act as an easy-to-read comprehensive guide \blue{that may help to facilitate} e-learning in the physics community. 

\section{Benefits, barriers, and key factors}
E-learning has a well-documented array of benefits and drawbacks. \cite{Zhang2004, Arkorful2015} Notably, it widens access to education and offers opportunities for pedagogical improvements by instructors. \cite{Mayadas2009} However, due to its drawbacks, e-learning is plagued by low-retention rates. \cite{Packham2004, Rosenthal2013, Murphy2017} 

In recent years, studies have identified some areas where e-learning may be advantageous for \textit{physics} teaching, especially through the use of smartphones, online learning systems, and social media\blue{.} \blue{E}ach of \blue{these technologies} can be implemented in the classroom for specific tasks---e.g. by facilitating group interaction and feedback loops, or by encouraging interest in coursework. \cite{Zhai2019, Zhu2012} By pairing these technologies with research-driven innovations, e-learning can be made more effective for physics education, as will be elaborated upon in the following sections. 

As previously mentioned, electronic instruction of physics faces specific barriers not found amongst teaching of the social sciences, humanities, and other natural sciences. First, e-learning is less \blue{well-suited} and less effective for science education that requires hands-on (e.g. laboratory) instruction. \cite{Arkorful2015} Additionally, students often find it difficult to visualize physical phenomena, especially those in 3-D (like the right-hand rule), through a screen. \cite{Kustusch2016} From a structure standpoint, teachers tend to have low \blue{competency} with technologies required for e-teaching physics. \cite{Badia2019} These factors, when paired with the lack of community students feel in online classes, contribute to higher withdrawal rates for online undergraduate introductory physics courses than in-person classes. \cite{Faulconer2018}

With these many costs and benefits in mind, it becomes clear that certain \blue{strategies for implementing} e-learning are key to its success. These \blue{best practices} are well-documented and extensively studied. \cite{Sun2008, Selim2007} The primary factors for successful and equitable e-learning include that:
\begin{enumerate}
	\item professional training is crucial and has been shown to improve teachers' acceptance of technology for physics instruction; \cite{Zhai2019, Zhai2018}
	\item real-time tech support is essential to successful instruction; \cite{Chow2017}
	\item participation in small-group collaborative learning correlates with deeper learning, \blue{increased} teamwork, and can increase students' sense of community; \cite{Brindley2009}
	\item mechanisms to directly combat high dropout rates for e-learners must be developed, including communication with lower-achieving students; \cite{Njenga2010} and
	\item inequalities should be considered when implementing e-learning, \blue{especially their} effect on \blue{access to} technology.
\end{enumerate}

\section{Addressing demographic concerns}
If e-learning is to be implemented equitably, economic concerns \blue{must be addressed}. The abrupt shift to an e-learning environment caused by COVID-19-induced closures generated significant difficulties for lower income families across the globe. These closures ``disproportionately affect vulnerable groups, in particular students with disabilities and those reliant on their educational institution for food, shelter, residency, and safety.'' \cite{Berger2020} Income-driven disparities are worse in urban centers like New York City, where 10\% of students were homeless or had unstable housing last year. \cite{VanLacker2020, NCHE2020} These cities are especially vulnerable to COVID resurgences due to high population densities. 

Data also shows that home computer availability in the U.S. scales with household income. \cite{Anderson2018} Alternatively, smartphone access is nearly ubiquitous amongst both teens and adults\blue{; it is} nearly uniform amongst people of varying gender, race, ethnicity, and socioeconomic status.\blue{\cite{PewMobile2019, Anderson2018}} Smartphones also have utility in addressing other demographic differences, such as empowering visually and hearing impaired students. \cite{Coca2017} The smartphone is evidently a key tool to \blue{address educational inequality and improve} physics education in the wake of COVID-19, and will be \blue{discussed} more thoroughly in the following section. 

It is well known that sex and gender inequality is rampant in the sciences, especially in physics. Although the percentage of female scientific authors increased substantially from 12\% in 1955 to 35\% in 2005, both physics and math still had female representations of 15\%. \cite{Huang2020} Additionally, within the classroom, female students have fewer successful learning and identity-forming experiences than males. \cite{Hazari2010}

Gender and socioeconomic differences can also be found in the use of technology for e-learning. \cite{Porter2006} Although it has been shown that there is no difference in scientific literacy across genders, males may show better performance in science practices because of their \blue{technological knowledge base} formed from daily activity. \cite{Pramuda2019} Multiple studies have shown \blue{that the success of technology is affected by} gender differences in \blue{its} perceived usefulness, perceived ease of use, and attitude towards \blue{its} use.\cite{Padilla-Melendez2013, Wong2012, Moran2010, Pynoo2011, Rasimah2011, Teo2011, Sumak2011} Educators should therefore communicate with their students to identify deficiencies in technological aptitude and comfort before electronic course instruction. They should utilize feedback loops integrated into learning management systems to continuously address the needs of underrepresented and disadvantaged students. These support structures must be designed and backed by the educational institutions, as teachers rarely have the resources or time to both develop their own feedback systems and implement them within the classroom. 

Lastly, when considering partial campus returns, household internet availability, technology access, and difficult home situations must be considered in selecting populations that will be allowed to return for on-campus instruction. These actions will assist \blue{in} making science education more accessible and \blue{will address} inequality prevalent in the sciences. 

\section{Smartphones as educational tools}

Over the last decade, cell phones have increasingly distracted students in the classroom. However, when teachers permit their use, smartphones can be effectively transformed into a \blue{learning tool}. They can facilitate the use of social media and learning management systems within and outside of the traditional classroom, and effectively complement other technologies. \cite{Ott2017} Researchers have advocated for smartphone use in teaching, arguing that smartphones offer benefits of ``rich content deliverability, knowledge sharing, and dynamic learning activities where students can expect to experience multiple channels of interactions in learning.''\cite{Buchholz2016, Anshari2017} Comprehensive lists of the advantages of smartphone use in the classroom are readily available.\cite{Coca2017, Attewell2005, Kolb2011, Duncan2012} \blue{S}uch advantages include \blue{the smartphone's} ability to encourage collaborative learning, students' existing familiarity with smartphones, and the fact that 96\% of young adults (aged 18--29) in the U.S. own smartphones.\cite{PewMobile2019} Additionally, if implemented properly, smartphone use can raise curiosity \blue{about physics} content while simultaneously \blue{introducing minimal distractions} and \blue{having} no dependence on gender, self-concept, or experimental experience.\cite{Sans2015, Hochberg2018} This section will elaborate on the physics-specific uses and advantages of smartphones for e-learning during the COVID-19 pandemic and beyond.

Smartphones are especially use\blue{ful} in laboratory settings due to their multitude of \blue{integrated} high-precision sensors and analysis tools. As described by Kolb, ``many teachers are discovering that a basic cell phone can be the Swiss army knife of digital learning tools.''\cite{Kolb2011} Such sensors include sound meters, accelerometers, magnetometers, proximeters, gyroscopes, photometers, cameras, GPS, and barometers. \cite{Vieyra2020} In order to access \blue{these sensors} directly, a host of physics toolbox and lab function apps have been developed.  \blue{Ideally, an app used for data collection and analysis should be free, easy to use, intuitive, open-source, and allow for processing and exportation of data as needed. \cite{Alexandros2020}} \blue{Some examples of apps} include the Physics Toolbox Sensor Suite, phyphox, Sensor Kinetics, Sensors Toolbox, and Sensors Pro. The former four are available as free apps (some with premium versions), while the latter is paid. 

Table I contains a list of smartphone-based lab experiments that can be used in undergraduate (and high school) introductory physics labs with little other equipment. These experiments are particularly relevant for planning post-COVID, on campus/small group learning, where smartphones can be employed to limit the use of shared equipment. Table II contains a similar list, but \blue{is} specifically geared towards smartphone-based lab experiments that can be conducted outside of the lab or at home. These experiments should be adapted \blue{ so that the lab instruction is aligned with AAPT Recommendations for the Undergraduate Physics Laboratory Curriculum}.\cite{Kozminski2014} \blue{One} process for implementing such a transition has been explicitly laid out.\cite{Holmes2019}

It has been shown that smartphone experiments ``may be more effective in improving students' understanding of acceleration with respect to traditional `cookbook' and real-time experiments,'' with the most significant improvements seen \blue{in} students' critical deductive thinking capability \blue{when} designing their own experiments.\cite{Mazzella2016} In order to carry out this approach, teachers are encouraged to adapt the POE method (predict-observe-explain) for smartphone-based experiments---have students predict the results of an experiment, collect data with smartphone sensors, and explain the resulting phenomenon theoretically. By allowing students to utilize this method in association with familiar smartphone technology, improvements in conceptual understanding of underlying phenomena can be achieved. 

The smartphone can be utilized in lecture sections as well. For example, the slow motion camera has been used \blue{to demonstrate} center of mass rotation, the Doppler effect, a frustrated Newton's cradle, the falling chimney effect, and tautochrones. \cite{Lincoln2017} The smartphone can be paired with external sensors like a thermal imaging camera to demonstrate phenomena such as work and energy transfer within the body. \cite{Kubsch2017} Additionally, pairing smartphones with a smart student response system can promote active physics learning in the classroom. \cite{Coca2017} These uses of smartphones will help transform it into a device with utility in the virtual classroom and laboratory. 

\begin{table}[h!]
\scriptsize
\centering
\caption{Smartphone-based lab experiments}
\begin{ruledtabular}
\def\arraystretch{0.7}
\begin{tabular}{c c}
Subject & Topic, Citation \\
\hline
Kinematics & Gravitational Acceleration  \cite{Pili2018a} \\
Dynamics & Static Friction  \cite{Kapucu2018} \\
Dynamics & Kinetic Friction \cite{Coban2019, Baldock2016} \\
Dynamics & Atwood Machine   \cite{Lopez2018} \\
Work \& Energy & Energy Conservation   \cite{Namchanthra2019, Pierratos2018} \\
Impulse \& Momentum & Collisions \cite{Vogt2014} \\
Impulse \& Momentum & Impulse   \cite{Ayop2017} \\
Impulse \& Momentum & Ballistic Pendula   \cite{Sanders2020} \\
Impulse \& Momentum & Collisions/Magnetism   \cite{Seeley2018} \\
Oscillations \& Waves & Spring Constants   \cite{Pili2018b, Pili2019} \\
Oscillations \& Waves & Simple/Damped Oscillations  \cite{Sans2013, CastroPalacio2013} \\
Oscillations \& Waves & Harmonic Series   \cite{Jaafar2016} \\
Oscillations \& Waves & Doppler Effect  \cite{GomezTejedor2014} \\
Rotation & Rotational Motion  \cite{Porn2016} \\
Rotation & Coriolis Acceleration  \cite{Shakur2016} \\
Rotation & Angular Acceleration/Spinning Discs   \cite{Gomes2018} \\
Rotation & Damped Rotational Motion   \cite{Klein2017} \\
Fluids \& Pressure & Fluid Mechanics  \cite{Goy2017} \\
Fluids \& Pressure & Surface Tension/Dispersion Relation  \cite{Wei2015} \\
Thermal Physics & Introductory Thermodynamics   \cite{Silva2018} \\
Electricity & Skin Depth Effect \cite{Rayner2017} \\
Electricity & Eddy Currents  \cite{Tomasel2012} \\
Magnetism & Faraday's Law   \cite{Soares2019} \\
Magnetism & Inductive Metal Detector  \cite{Sobral2018} \\
Magnetism & Basic Magnetism \cite{Arribas2015, Ogawara2017} \\
Magnetism & Collisions/Magnetism   \cite{Seeley2018} \\
Magnetism & Eddy Currents   \cite{Tomasel2012} \\
Light & Malus' Law   \cite{Rosi2020, Monteiro2017} \\
Light & Absorption / Scattering  \cite{Malisorn2019} \\
Light & Lens Equation  \cite{Freeland2020} \\
Light & Brewster's Angle  \cite{Chiang2019} \\
Light & Linear Light Source  \cite{Salinas2018} \\
Light & Properties of EM Waves \cite{Onorato2020} \\
Quantum Mechanics & Double-Slit \cite{Ghalila2018} \\
Quantum Mechanics & e/m Experiment  \cite{Pirbhai2019} \\
Astronomy & Astronomy \& Seasons  \cite{Durelle2017} \\
Materials Physics & Polymer Physics  \cite{Vandermarliere2016} \\
Electronics & Oscilloscopes  \cite{Forinash2012} \\
\end{tabular}
\end{ruledtabular}
\label{bosons}
\end{table}

\begin{table}[h!]
\scriptsize
\centering
\caption{Smartphone-based at-home experiments}
\begin{ruledtabular}
\def\arraystretch{0.7}
\begin{tabular}{c c}
Subject & Topic, Citation \\
\hline
Kinematics & Gravitational Acceleration \cite{Schwarz2013} \\
Kinematics & Free Fall \cite{Vogt2012, Kim2020, Kuhn2016} \\
Kinematics & Basic Kinematics \cite{Testoni2016} \\
Dynamics & Air Resistance \cite{Azhikannickal2019} \\
Dynamics & Drag Coefficient \cite{Fahsl2018} \\
Impulse \& Momentum & Collisions \cite{deJesus2016} \\
Impulse \& Momentum & Conservation of Momentum \cite{Pereira2017} \\
Oscillations \& Waves & Acoustics \cite{Kuhn2013} \\
Oscillations \& Waves & Pendula \cite{Kuhn2012} \\
Oscillations \& Waves & Speed of Sound \cite{Hellesund2019, Parolin2013, Yavuz2015, Staacks2019} \\
Oscillations \& Waves & Mechanical Wave Physics \cite{Bonato2017} \\ 
Oscillations \& Waves & Acoustic Resonance \cite{Monteiro2018} \\
Oscillations \& Waves & Sound Directivity \cite{Hawley2018} \\
Oscillations \& Waves & Acoustic Modeling \cite{Thees2017} \\
Oscillations \& Waves & Pressure Waves \cite{Muller2016} \\
Oscillations \& Waves & Hooke's Law \cite{Smith2019} \\
Rotation & Rolling Motion \cite{Dilek2019} \\
Rotation & Radial Acceleration \cite{Vogt2013} \\
Rotation & Phase Space \cite{Monteiro2014} \\ 
Rotation & Parallel Axis Theorem \cite{Salinas2019} \\
Rotation & Angular Velocity \cite{Pili2018c} \\
Rotation & Mechanics \cite{Chevrier2013} \\
Rotation & Centripetal Acceleration \cite{Mau2016} \\
Fluids \& Pressure & Stevin's Law \cite{Macchia2016} \\
Fluids \& Pressure & Atmospheric Pressure Profiles \cite{Monteiro2016} \\
Fluids \& Pressure & Fluid Dynamics \cite{Smith2019} \\
Light & Ray Optics \cite{Girot2020} \\
Special Relativity & Time Dilation \cite{Underwood2016} \\
Nuclear \& Particle Physics & Radiation \cite{Kuhn2014} \\
Astronomy & Orbital Angular Velocity \cite{Meisner2014} \\
\end{tabular}
\end{ruledtabular}
\label{bosons}
\end{table}

\section{Integrating social media in ``the classroom''}
The \blue{pervasiveness of} social media (SM) presents an intriguing opportunity \blue{for students to collaborate on} physics inside and outside of the traditional classroom. SM comprises an array of online tools through which users can quickly create and share content digitally---e.g. Twitter, Facebook, YouTube, and Wikipedia. The integration of SM within the learning environment \blue{offers} students self-agency \blue{in} their learning and career planning. Despite these positive \blue{attributes}, most universities continue to rely on more conservative, established learning management systems and environments. \blue{Ignoring social media} prevents educators from capitalizing on \blue{the collaborative potential of social networks} and \blue{the} associated social skills that students bring into the classroom.\cite{McLoughlin2010} \blue{Concern due to privacy laws is a primary reason that educators are relucant to use SM;} \blue{t}o comply with these laws while using SM, instructors should never share grades, records, or personal information via platforms not maintained by the university. Furthermore, faculty who use SM for e-learning should discuss and include a statement in their syllabi about proper conduct and expectations for online privacy, \blue{and should also} consult their university SM/privacy guidelines.\cite{Rodriguez2011, Joosten2012} Other reservations held by faculty about the implementation of SM in the classroom include the following:\cite{Manca2016, Bouhnik2014, Hasiloglu2020, Moran2012, Buchanan2013, Cao2013, Anderson2018}
\begin{enumerate}
	\item Not all students have smartphone access (although, in the U.S., $\sim$95\% do).
	\item Cultural/Social---\blue{Instructors} show reluctance because: 
	\begin{itemize}
		\item there is a perceived erosion of traditional roles and difficulties in managing relationships with students;
		\item \blue{students may engage in inappropriate chatting}; and
		\item language barriers \blue{and unconscious biases} \blue{can lead to misunderstandings}.
	\end{itemize}
	\item Pedagogical---\blue{Perceived usefulness is an important motivator for technology usage, but instructors often rate SM poorly in this category.  Many instructors perceive direct, face-to-face relations with students as indispensable and more effective than SM use.}
	\item Administrative/Institutional---Instructors show reluctance because \blue{the} success \blue{of teaching technology} is reliant on financial investment and institutional support provided by the university (elaborated in final section).
\end{enumerate}

Not all instructors hesitate to employ SM in the classroom, and its uses vary by field. In a survey of 459 secondary teachers, almost all teachers used SM in the class. There were\blue{, however} some differences in the modes of use; for example, teachers in the natural sciences used SM less often for the facilitation of self-regulated learning.\cite{Matzat2015} Conversely, another study showed that university faculty used SM less (41\% \blue{of faculty use} at least one tool on a monthly basis). Younger faculty used SM more than their colleagues, particularly Twitter---though it was concluded that age differences require further investigation. Math, computer science, and natural science faculty used SM less than those in humanities and social sciences. \cite{Manca2016} The tendency for natural science faculty to use SM less than their colleagues is attributed to ``a lack of relevant content on social media sites for their particular discipline.'' \cite{Moran2012} The dearth of relevant content has been explained by a trend in faculty consuming rather than producing digital resources (which requires a large time investment). \cite{Hargittai2008} Science faculty, as a result, tend to prefer blogs/Wikipedia and Youtube/Vimeo information sources to promote collaborative learning, rather than Facebook/Twitter type communication SM channels. \cite{Manca2016}

\blue{T}o demonstrate the effectiveness of SM within the classroom and to address the barriers presented previously, \blue{some specific examples can be offered}. One principle of high-impact online education is faculty/teaching assistants providing timely feedback to students outside of class.\blue{\cite{Bao2020, Bouhnik2014}} This task can be assisted through SM communication channels\blue{.} \blue{F}or example, WhatsApp can facilitate student-teacher interaction within online college courses. \cite{Amry2014} Connecting with students outside of classroom hours through WhatsApp can permit physics teachers to identify problems that are not recognized during the traditional class hours.\cite{Klieger2019} Other similar messaging apps including Slack, Discord, GroupMe, and Google Hangouts can replace WhatsApp with similar functionality.  Overall, SM helps teachers share information, questions, and insights to promote curiosity in physics. \cite{Hasiloglu2020}

Perhaps of most importance, \blue{a} lack of community is often blamed for the high withdrawal rates of online learning. \cite{Hart2012} Microblogging (e.g. Twitter) has been shown to combat this flaw, strengthening \blue a sense of community in virtual classes within higher education. \cite{Hsu2011} Classroom-specific Twitter threads can be used to provide \blue{course} updates and facilitate academic conversations in a manner familiar to students. \cite{Page2015, Burden2014} WhatsApp can be employed \blue{to encourage} student-student messaging and sharing of ideas. \cite{Amry2014} Similarly, the use of Facebook groups for sharing ideas and support, asking questions, and participating in discussions has been shown to promote a virtual student learning community. \cite{Schoper2017} \blue{Therefore,} integration of SM\blue{---}especially through inclusive technologies such as the smartphone\blue{---}can be key to battling low retention rates in virtual education during the COVID-19-induced closures.

Research on innovative practices is crucial for adapting to changing learning environments. \blue{S}haring of effective practices can assist in the re-thinking of pedagogies\blue{, and }could shift attitudes from resistance to a welcomeness in using SM to assist and improve physics teaching in higher education. 

\section{Other Novel At-Home Techniques}
Smartphone use \blue{is} a promising way to do physics at home, but other technologies can be used in complement with smartphones or \blue{can} replace them for various e-learning tasks when they are unsuitable. For example, experimental kits provide students the opportunity to conduct physics right on the kitchen table, and can be instructor-provided or student-assembled.  Such kits have been implemented in the classroom, \cite{Gibney2016} for massive open online courses (MOOCs), \cite{DeBoer2019} in open universities, \cite{Kennepohl2017} for in-class demonstrations, \cite{Stpanka2015} and for experimental distance learning. \cite{Long2012, Hendricks2009} Kits can even be paired with smartphones as data collection and analysis devices to increase student comfort with the experiments. 

Given \blue{the inaccessibility of} physical laboratory equipment, experiments can also be conducted remotely. Virtual and remote labs have been around since commercialized internet became prevalent across the world, and their use has expanded significantly over time. \cite{Heradio2016, Gomes2009} These are \textit{real} experiments \blue{(}housed at hosting institutions\blue{)} \blue{which are} accessed and controlled by individual users through the internet.\cite{Ma2006, Matarrita2016} One such facility is FARLabs, led by La Trobe University, which allows users to remotely access lab technologies for real-time experiments. \cite{Hoxley2014}  Some researchers are endeavoring to promote remote labs through sharing economy platforms such as LabsLand. \cite{Orduna2016} Remote experiments can be used to teach many aspects of physics, for example, radioactivity \cite{Jona2013} and electronics. \cite{Tawfik2012} Such a platform provides clear financial advantages over physical analogs\blue{.} \blue{These platforms also allow for} better student access to equipment, \blue{increased} scheduling flexibility, a wider range of possible assignments and activities, and more opportunities for student-student collaboration.\cite{Zubia2012}

For instructional demonstrations of concepts and simple experiments simulations can be extremely useful. Simulations have been used in the physics classroom for many years, \cite{Jimoyiannis2001} and much research has been conducted on successful approaches for their use. \cite{Wieman2008} Two such simulation bases are the PhET project developed by the University of Colorado and PhysClips of the University of New South Wales. \cite{Wieman2010, Hatsidimitris2012} Many simulations and other resources can also be found on The Physics Source at AAPT's ComPADRE site. \cite{MacIsaac2006} Simulations can be especially advantageous for instructors struggling with the extra preparation time required for online courses.

Lastly, \blue{free online materials can be useful and are often overlooked.}\cite{Thompson2011} Whereas YouTube videos---such as those generated by Physics Girl, minutephysics, etc.---might be used by students intermittently, the consistent use of resources such as Khan Academy and HyperPhysics can fill gaps in student knowledge, or act as a support system for a \blue{struggling student}.\cite{Lindstrom2015} Lists of similar online resources can be found readily.\cite{Kettle2013} 

\section{Recommendations for institutional administrators}
Whereas this manuscript is \blue{aimed at assisting physics educators shift to online learning}, effective recommendations for implementing e-learning necessarily include an administrative component. \cite{Toquero2020} The effectiveness of pandemic-induced e-learning will depend on educational institutions realigning with and embracing the necessary structural changes associated with it. \cite{Njenga2010, Peppard2004} Such changes are outlined below.

\begin{enumerate}
	\item 
	Online mental health and medical services should be expanded. \cite{Sahu2020} It was found that 20--35\% of the 2,530 surveyed students and workers at a Spanish university reported moderate to severe symptoms of anxiety, depression, and stress after COVID-19 school closures. \cite{Gonzalez2020} Similarly, nearly half (46\%) of Australian young people studying at home are ``vulnerable to adverse effects on their educational outcomes, nutrition, physical movement, social, and emotional wellbeing.'' \cite{Brown2020} Universities (and other educational institutions) are recommended to:
	\begin{itemize}
		\item expand the availability of online counseling services; and
		\item encourage faculty to employ technology as a means to increase interactivity, enrich learning, and enhance the student experience, .
	\end{itemize}
	\item
	Direct financial investment into e-learning should be a priority. \cite{Njenga2010} An analysis of blended learning at one university showed that student satisfaction was best predicted by the availability of university resources. \cite{Calderon2012} Another study showed that the top faculty-identified needs for successful e-learning are multimedia development support and real-time help desks. \cite{Chow2017} Universities are recommended to:
	\begin{itemize}
		\item make expenditures related to internet access necessary for hybrid approaches; \cite{Guri-Rosenblit2006, MacKeogh2001}
		\item engage in hiring or contracting of support staff for IT; \cite{Chow2017} and
		\item ensure proper compensation for instructors; (\blue{this is} important for quality online instruction \cite{Mihhailova2006}).
	\end{itemize}
	\item
	Pedagogical research, data collection, and evidence-based practices focused on e-learning should be expanded. Student feedback can be motivated by effective communication mechanisms integrated into a student's online learning space, \cite{Hatziapostolou2010, Jara2010} and has been shown to be of great value in improving blended course quality. \cite{Calderon2012} Universities are recommended to:
	\begin{itemize}
		\item \blue{organize a system to analyze feedback data, identiffy problem points, delegate responsibility for addressing them, and report back to the students on resulting actions.\cite{Watson2003}} \blue{I}ntegrat\blue{ing} easy-access course feedback into virtual learning management system \blue{is an excellent way to} ``close the feedback loop''; and
		\item adopt a hiring and promotion process that factors in teaching achievement through student feedback. This will help incentivize research-driven innovation and teaching practices in the classroom. 
	\end{itemize}
	\item
	Teacher training capabilities for multiple modes of e-learning should be expanded. \cite{Njenga2010} Training is essential to the effective delivery of electronic physics instruction, \cite{Afzal2015} and has been demonstrated to lead to instructors' enthusiastic acceptance of mobile technology for teaching.\cite{Zhai2018, Zhai2019}
\end{enumerate}

\begin{figure} [h!]
	\centering
	\includegraphics[width=\linewidth]{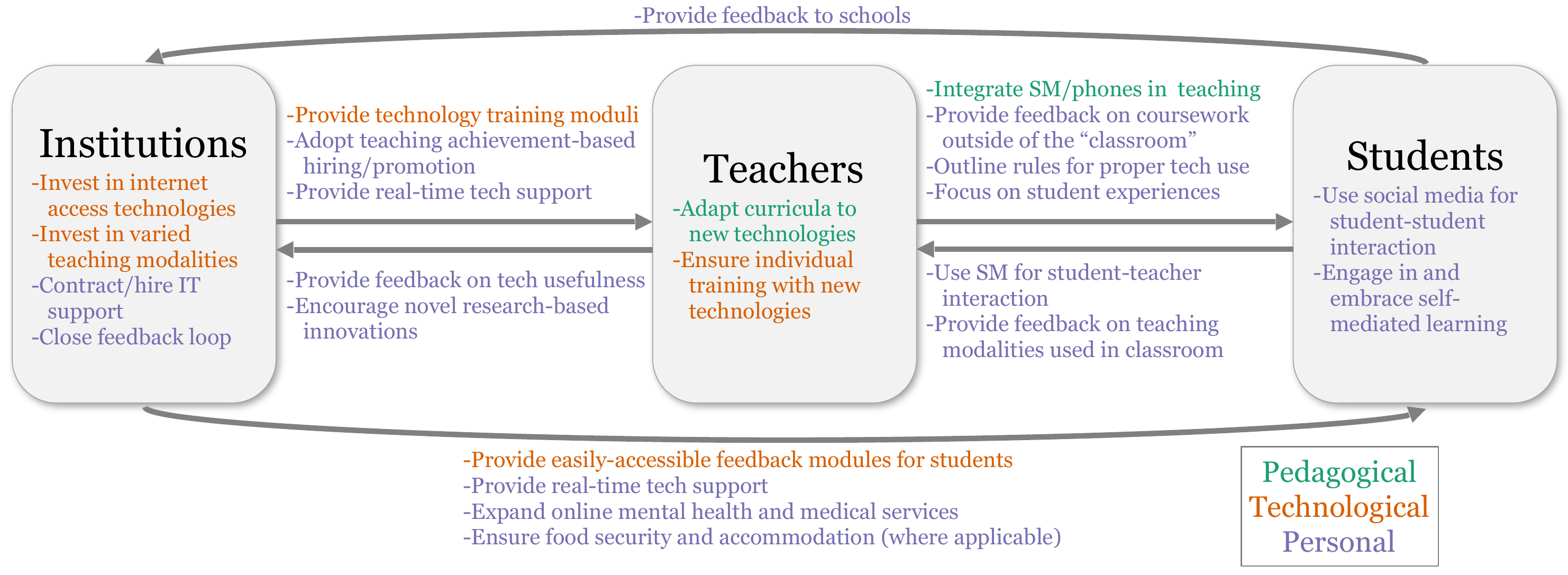}
	\caption{A graphical representation of support mechanisms for improving e-teaching and e-learning of physics at the university level (color online)}
	\label{fig1}
\end{figure}

A general outline of the contributions necessary from administrators, faculty, and students is offered in Fig. \ref{fig1}. As learning institutions resume education during COVID-19, \blue{f}aculty should encourage \blue{administrators to adopt} these recommendations \blue{as they are essential to the} success of education under circumstances induced by the pandemic. 

\section{Conclusions}
The COVID-19 pandemic thrust learners and educators across the world into a new environment, in which e-learning became the foremost method of education. As the community is unsure about how this pandemic will persist, it is of paramount importance to embrace e-learning in physics education. First, demographic concerns were addressed, including technology's association with income and gender differences in physics. Consideration of demographics is key to the equitable implementation of e-learning.  \blue{Second,} \blue{i}t was proposed that adopting research-driven innovation will help teachers adapt curricula to the changing needs of students in the wake of the pandemic. The smartphone was explored as an educational tool\blue{;} \blue{its advantages} in the classroom and its range of sensors and apps for use in the laboratory \blue{were identified}. Nearly 80 examples of smartphone-based lab and at-home introductory physics experiments were \blue{provided} and sorted by subject. Following \blue{that}, a guide for the use of social media as a classroom tool was presented. While smartphones and social media are key for some aspects of e-learning, other technologies like remote labs and experimental kits can complement their use effectively. Lastly, a guide for institutional administrators \blue{was offered.  This guide} highlight\blue{ed} the need for online mental health/medical services, financial investment in e-learning, pedagogical research initiatives, and teacher training.  This manuscript should be utilized by the physics community as a whole to help guide the \blue{implementation} of fruitful electronic learning practices. 

\begin{acknowledgments}

Much thanks to A. J. Nijdam, P. Johnson, and L. Doughty (Georgetown), S. Fisher (William \& Mary), and A. Sedlack (NIH) for careful review of the manuscript, as well as my advisor, M. Paranjape, for encouraging me in undertaking this project. Thank you to both the NSF for current funding through award number CBET-1938995, the Achievement Rewards for College Scientists-Metro Washington Chapter, and the Forster Family Foundation for general research funding.

\end{acknowledgments}

\end{document}